# Femtosecond photocurrents by the Dresselhaus bulk spin-galvanic effect in an inversion-asymmetric ferromagnet

Junwei Tong[1,*,+], Zdenek Kaspar[1,2,3*], Afnan Alostaz[1,4], Reza Rouzegar[1], Chihun In[1], Tim Titze[5], Maximilian Staabs[5], Genaro Bierhance[1], Yanzhao Wu[6], Holger Grisk[7], Jakob Walowski[7], Markus Münzenberg[7], Felicitas Gerhard[8], Johannes Kleinlein[8], Tobias Kießling[8], Charles Gould[8], Laurens W. Molenkamp[8], Xianmin Zhang[6,+], Daniel Steil[5], Tom S. Seifert[1,+], Tobias Kampfrath[1]

[1]Department of Physics, Freie Universität Berlin, Berlin, Germany

[2]Institute of Physics, Czech Academy of Sciences, Prague, Czech Republic

[3]Faculty of Mathematics and Physics, Charles University, Prague, Czech Republic

[4]Peter Grünberg Institut-6, Forschungszentrum Jülich GmbH, Jülich, Germany

[5]I. Physikalisches Institut, Georg-August-Universität Göttingen, Göttingen, Germany

[6]School of Material Science and Engineering, Northeastern University, Shenyang, China

[7]Institute of Physics, Greifswald University, Greifswald, Germany

[8]Physikalisches Institut der Universität Würzburg (EP3) and Institute for Topological Insulators, Würzburg, Germany

*Contributed equally
+Corresponding authors.

**Abstract:** We study ultrafast photocurrents in thin films of a model ferromagnetic metal with broken bulk inversion symmetry, the half-metallic Heusler compound NiMnSb, following excitation with an optical pump pulse with photon energy 1.55 eV. Remarkably, in terms of the direction of the sample magnetization $\boldsymbol{M}$, all photocurrents are found to be a superposition of a component with Rashba- and Dresselhaus-type symmetry. We explain the Dresselhaus bulk photocurrent as follows: Pump-induced electron heating induces an excess of spin $\boldsymbol{\mu}_s \parallel \boldsymbol{M}$, which transfers spin angular momentum into states with Dresselhaus-type spin-momentum locking. The resulting charge current relaxes on a time scale of 10 fs by momentum relaxation and, thus, follows $\boldsymbol{\mu}_s$ quasi-instantaneously. The relaxation of $\boldsymbol{\mu}_s$ is governed by the cooling of the electrons and not by the significantly slower spin-lattice relaxation of half-metals. Our findings add the Dresselhaus spin-galvanic effect (SGE) to the set of ultrafast spin-charge-conversion phenomena. They indicate a route to more efficient spintronic terahertz emitters and detectors based on the volume scaling of the bulk SGE.

**Figures**

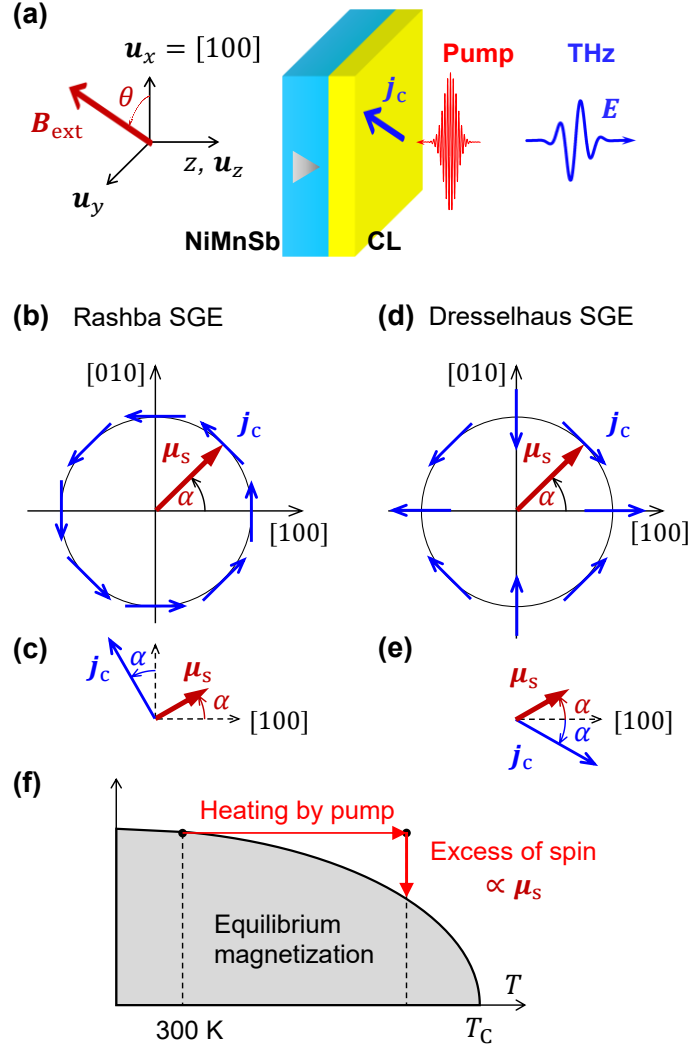

**Fig. 1.** Concept of the experiment. **(a)** A femtosecond optical pump pulse excites a ferromagnetic NiMnSb|CL thin-film stack made of the ferromagnetic Heusler compound and a capping layer CL of Ru or MgO. The resulting ultrafast photocurrent with density $\boldsymbol{j}_\mathrm{c}$ emits a THz electromagnetic pulse with electric field $\boldsymbol{E}$ behind the sample [Eq. (2)]. The unit vectors $\boldsymbol{u}_x$, $\boldsymbol{u}_y$, $\boldsymbol{u}_z$ indicate the $x$, $y$, $z$ axes, where $\boldsymbol{u}_x$ is parallel to the NiMnSb [100] axis. The in-plane magnetization $\boldsymbol{M}$ of NiMnSb is set by an external magnetic field $\boldsymbol{B}_\mathrm{ext}$ at an angle $\theta = \sphericalangle(\boldsymbol{B}_\mathrm{ext}, [100])$. The grey triangle indicates the broken bulk inversion symmetry of NiMnSb. **(b)** Relationship between the excess of spin $\boldsymbol{\mu}_\mathrm{s}$ (red arrow) and induced charge-current density $\boldsymbol{j}_\mathrm{c}$ (blue arrow) for a SGE with Rashba symmetry. **(c)** The angle of $\boldsymbol{j}_\mathrm{c}$ and $\boldsymbol{\mu}_\mathrm{s}$ relative to the [100] axis is $\alpha$ and $\alpha + 90°$, respectively. **(d)** Same as panel (b), but for the SGE component with Dresselhaus symmetry. **(e)** The angle of $\boldsymbol{j}_\mathrm{c}$ and $\boldsymbol{\mu}_\mathrm{s}$ relative to the [100] axis is $+\alpha$ and $-\alpha$, respectively. **(f)** In our experiment, the pump pulse induces $\boldsymbol{\mu}_\mathrm{s}$ by ultrafast electrons heating. The resulting difference between the instantaneous magnetization and the equilibrium magnetization at the elevated temperature scales with $\boldsymbol{\mu}_\mathrm{s}$.

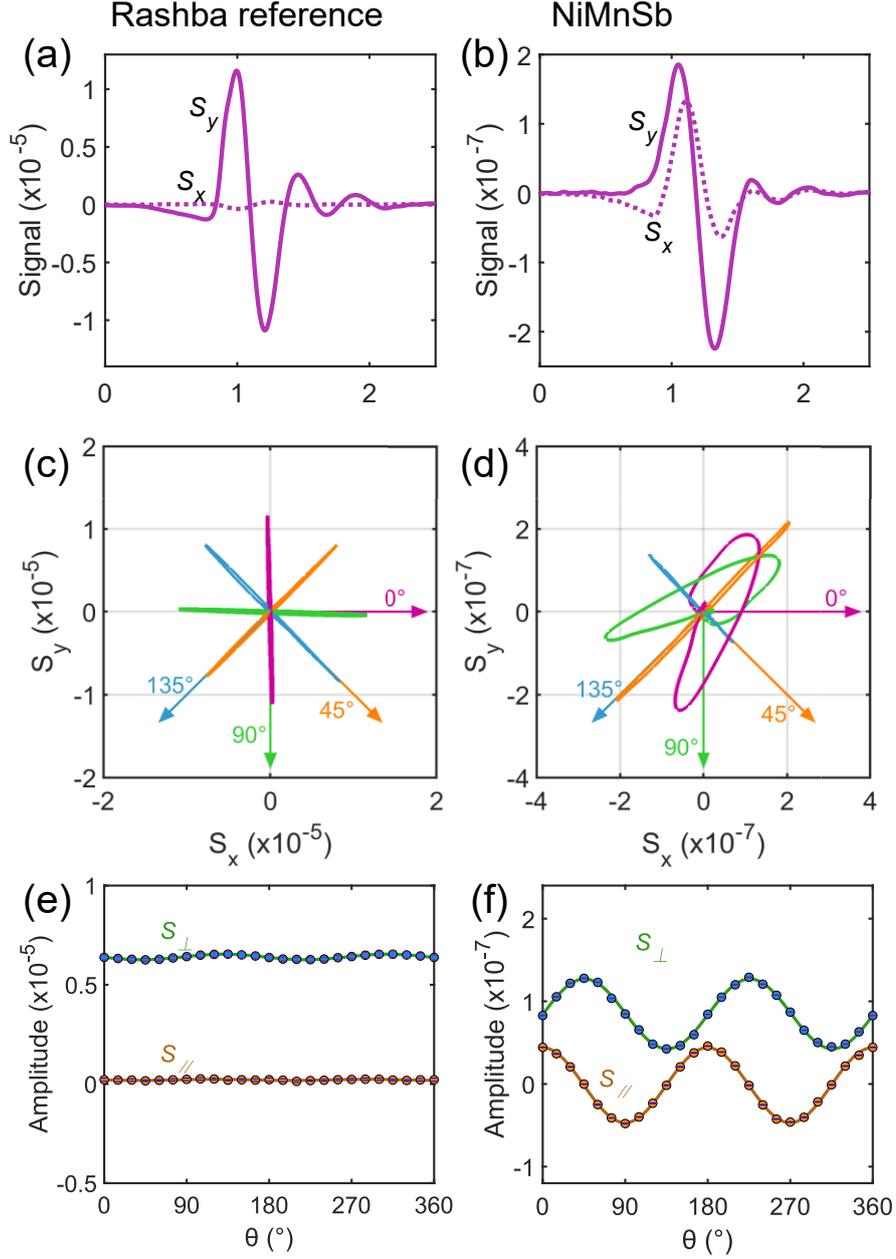

**Fig. 2.** Typical THz-emission raw data. **(a)** Terahertz electro-optic signals $S_x(t)$ and $S_y(t)$ from the Rashba reference and **(b)** NiMnSb sample. Here, the angle $\theta = \sphericalangle(\boldsymbol{B}_{\text{ext}}, [100])$ between the external field $\boldsymbol{B}_{\text{ext}}$ and the $x$ and, thus, [100] axis of NiMnSb is set to 0° [Fig. 1(a)]. **(c)** Time traces $\boldsymbol{S}(t) = (S_x, S_y)(t)$ of THz signals from the Rashba reference for 4 directions $\theta = 0°, 45°, 90°, 135°$ of $\boldsymbol{B}_{\text{ext}}$, which are indicated by arrows. **(d)** Same as (c), but for NiMnSb. **(e)** Amplitudes of the projection of the THz signal $\boldsymbol{S}(t)$ on the directions parallel ($S_{\parallel} = \boldsymbol{S} \cdot \boldsymbol{u}_{\theta}$) and perpendicular ($S_{\perp} = \boldsymbol{S} \cdot \boldsymbol{u}_{\theta+90°}$) to $\boldsymbol{B}_{\text{ext}}$ for the Rashba reference sample and **(f)** NiMnSb. The signals in panels (a)-(d) were measured with a 1 mm thick ZnTe(110) detector whereas for panels (e)-(f) 250 μm thick GaP(110) was used.

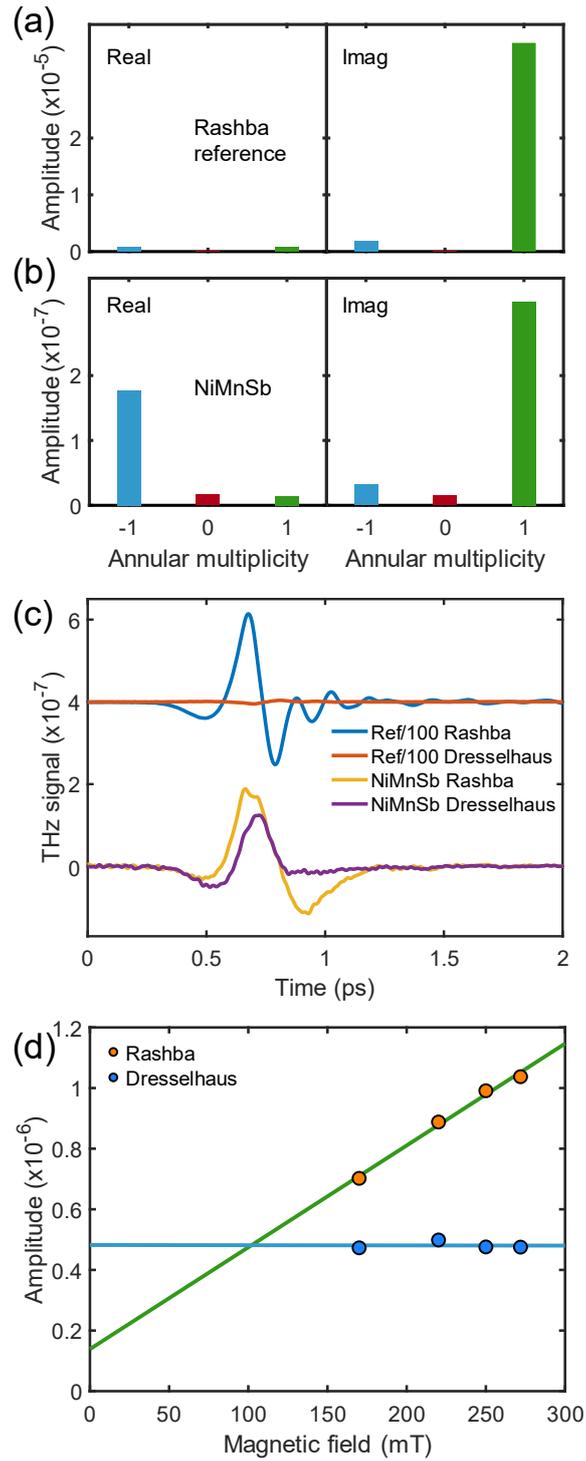

**Fig. 3.** Photocurrent characteristics. **(a)** Real and imaginary part of the amplitudes of the THz signals from the Rashba reference sample following Fourier transformation with respect to $\theta$. **(b)** Same as in panel (a), but for NiMnSb. **(c)** THz-signal components with Rashba and Dresselhaus symmetry from Rashba reference sample (Ref) and NiMnSb sample. **(d)** Amplitude of the THz-signal components from NiMnSb with Rashba- and Dresselhaus-like symmetry as a function of $|\boldsymbol{B}_{\text{ext}}|$.

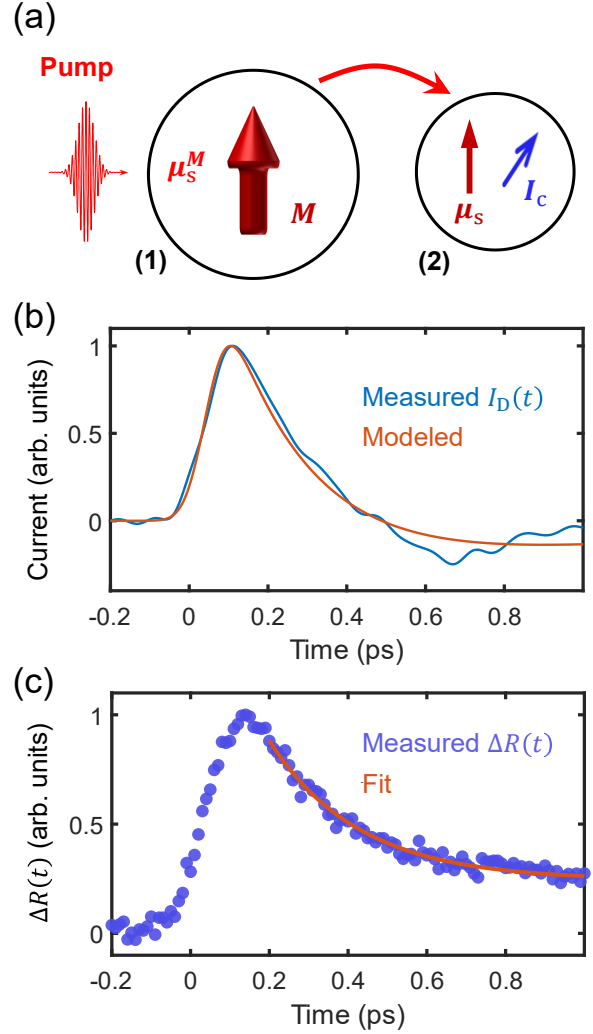

**Fig. 4.** Origin of the Dresselhaus photocurrent. **(a)** In the proposed scenario, NiMnSb is considered to contain an electron subsystem (1) with localized magnetic moments of magnetization $\boldsymbol{M}$ and an electron subsystem (2) that provides the SGE. The optical pump pulse induces an excess of spin $\boldsymbol{\mu}_s^M$ in (1) [Fig. 1(f)], which drives spin transfer to (2). Consequently, a spin excess $\boldsymbol{\mu}_s$ builds up in (2) and gives rise to a charge current $\boldsymbol{I}_c$ by the SGE [Eq. (1)]. Because NiMnSb is a half-metal, $\boldsymbol{\mu}_s^M$ does not relax by spin transfer to the crystal lattice but rather by cooling of the hot electrons by the crystal lattice. **(b)** Extracted pump-induced charge-current sheet density $I_D(t)$ of NiMnSb with Dresselhaus symmetry. The red line is a fit by Eq. (17). **(c)** Pump-induced change in the sample reflectance at a probe wavelength of 515 nm. The red line is a fit by Eq. (16). Note the similar relaxation dynamics of (a) $I_D(t)$ and (b) $\Delta R(t)$.

**Motivation.**

Spin-to-charge conversion (SCC) and its inverse effect play a crucial role in spintronic devices based on thin-film stacks [Fig. 1(a)]. Examples of elementary operations include spin-current generation and detection [1], spin accumulation, and torquing of magnetization [2]. SCC has various manifestations: In centrosymmetric media, the inverse spin Hall effect (ISHE) results in the conversion of a spin current into a charge current perpendicular to the polarization and propagation direction of the spin angular momentum. Importantly, in systems with broken inversion asymmetry, e.g., at interfaces or in each bulk unit cell, the spin galvanic effect (SGE), also known as inverse Rashba-Edelstein effect, makes an additional contribution to SCC [3]. It converts an excess $\boldsymbol{\mu}_\mathrm{s}$ of spin angular momentum into a charge current. The local total charge-current density $\boldsymbol{j}_\mathrm{c}$ is, thus, the sum

$$\boldsymbol{j}_\mathrm{c} = \gamma : j_\mathrm{s}\boldsymbol{\sigma} \otimes \boldsymbol{u}_z + \chi : \boldsymbol{\mu}_\mathrm{s}. \tag{1}$$

Here, $j_\mathrm{s}$ is the amplitude of the spin-current density $j_\mathrm{s}\boldsymbol{\sigma} \otimes \boldsymbol{u}_z$ that has spin polarization $\boldsymbol{\sigma}$ and flows along the $z$-axis normal to the thin-film stack [Fig. 1(a)]. The structure of the ISHE tensor $\gamma$ and the SGE tensor $\chi$ is dictated by the spatial symmetries of the system.

Macroscopically, in model materials with broken inversion symmetry, such as BiTeI, GaAs and NiMnSb, the SGE has two different contributions that possess Rashba-type [Fig. 1(b,c)] vs Dresselhaus-type symmetry [Fig. 1(d,e)]. The Rashba component of the induced charge current $\boldsymbol{j}_\mathrm{c}$ is orthogonal to the spin accumulation $\boldsymbol{\mu}_\mathrm{s}$ [Fig. 1(c)]. In contrast, for the Dresselhaus component, the $\boldsymbol{j}_\mathrm{c}$ direction is a mirror image of the $\boldsymbol{\mu}_\mathrm{s}$ direction at a certain crystal plane [Fig. 1(e)]. Microscopically, the SGE is often explained by locking of the momentum (wavevector) and spin of each electronic Bloch state. In high-symmetry situations, the resulting spin textures in momentum space are analogous to Fig. 1(b-e), with $\boldsymbol{j}_\mathrm{c}$ and $\boldsymbol{\mu}_\mathrm{s}$ substituted by momentum and spin [2].

To measure Rashba and Dresselhaus SGE or its inverse, experiments with optical, electrical or spin-precession excitation were used [4–8]. To gain further insights into the elementary mechanisms of SCC, it is important to consider them on the natural time scale of electron dynamics in solids, i.e., the femtosecond scale [9]. This approach was already successfully applied to the ISHE [10–12] and the SGE with Rashba symmetry [13–16]. However, ultrafast signatures of the SGE with Dresselhaus symmetry have not yet been reported, despite their high relevance for spintronic functionalities, such as THz-field generation [10–12], and detection [17], as well as ultrafast spin control [18–20].

**In this article,** we study femtosecond photocurrents in thin films of the ferromagnetic half-metallic Heusler compound NiMnSb by THz-emission spectroscopy. Following excitation by an optical femtosecond pump pulse, we observe photocurrents that result from components with Rashba- and Dresselhaus-like symmetry. While the Rashba-like contribution predominantly arises from the ordinary Hall effect, the Dresselhaus-type component is explained by a pump-induced spin accumulation and its conversion into a charge current by the SGE in the NiMnSb bulk. The spin accumulation relaxes approximately with the time constant of the pump-induced change in the electron temperature, consistent with the slow spin-lattice relaxation of half-metals. Our results are relevant for the understanding of the bulk SGE on its natural time scale, spin dynamics in half-metals as well as the development of more efficient spintronic emitters and detectors of broadband THz electromagnetic pulses.

**Experiment.**

To address Eq. (1) at THz frequencies, we choose thin films of the metallic Heusler compound NiMnSb for the following reasons. First, NiMnSb has a bulk Dresselhaus contribution to SCC due to broken inversion symmetry [4]. Second, NiMnSb is a ferromagnetic metal (Curie temperature 730 K) [4,21–25] and, thus, allows us to induce a spin excess or spin accumulation $\boldsymbol{\mu}_s$ parallel to the NiMnSb magnetization $\boldsymbol{M}$ by ultrafast optical excitation [Fig. 1(f)]: As shown for a number of metallic ferromagnets [26,27], an optical femtosecond laser pulse rapidly heats the electronic subsystem and, thus, induces an excess of electron spin. This spin accumulation needs to be released to adapt the magnetization $\boldsymbol{M}$ to the elevated electron temperature. The ultrafast charge current $\boldsymbol{j}_c$ resulting from the SGE [Eq. (1)] emits an electromagnetic pulse with frequencies extending into the THz range [10,28–38]. Directly behind the sample [Fig. 1(a)], the THz electric field $\boldsymbol{E}(t)$ vs time $t$ is directly proportional to the current sheet density

$$\boldsymbol{I}_c(t) = \int \mathrm{d}z\, \boldsymbol{j}_c(z,t). \quad (2)$$

***THz setup.*** In our experiment, we excite the sample from the metal side with pump pulses (energy 1.6 nJ, photon energy 1.55 eV, nominal duration 10 fs) from a Ti:sapphire oscillator (repetition rate 80 MHz) under normal incidence [Fig. 1(a)]. The pump-induced THz electromagnetic pulse that propagates into the specular reflection direction traverses a pair of polarizers and is subsequently monitored by electro-optic sampling in a GaP(110) crystal (thickness 250 μm) at 1-7 THz and ZnTe(110) crystal (1 mm) at 1-3 THz [10]. The resulting signals $S_x(t)$ and $S_y(t)$ are related to field components $E_x(t) = \boldsymbol{u}_x \cdot \boldsymbol{E}(t)$ and $E_y(t) = \boldsymbol{u}_y \cdot \boldsymbol{E}(t)$ by transfer functions that can be measured. The $x$, $y$ and $z$ axes are permanently locked to the crystal axes of the fixed sample [Fig. 1(a)]. We focus on signals odd with respect to $\pm \boldsymbol{B}_{\mathrm{ext}}$. Here, $\boldsymbol{B}_{\mathrm{ext}}$ is an external magnetic field from a rotatable pair of permanent magnets that sets the direction of the sample magnetization $\boldsymbol{M}$. All beams are normally incident onto the sample surface.

***Samples and characterization.*** Our NiMnSb thin films are grown on a 200 nm-thick In$_{0.53}$Ga$_{0.47}$As buffer layer on an insulating substrate of Fe:InP by molecular-beam epitaxy and subsequently capped with a cap layer (CL) of Ru(2 nm) or MgO(5 nm) (see Supplemental Note 1). The resulting stack is Fe:InP||In$_{0.53}$Ga$_{0.47}$As(200 nm)|NiMnSb(10 nm)|CL. The sample sheet conductance is 3.6 MS/m. Measurements of the magneto-optic Kerr effect (MOKE) as a function of $\boldsymbol{B}_{\mathrm{ext}}$ yield a coercive magnetic field below 10 mT and a saturation magnetic field of 50 mT (Supplemental Note 2). In the THz experiment, the external magnetic field has a significantly larger amplitude of $|\boldsymbol{B}_{\text{ext}}| = 270$ mT and ensures that $\boldsymbol{M} \parallel \boldsymbol{B}_{\text{ext}}$.

As reference sample, we use the metal stack sapphire||W(2 nm)|CoFeB(1.8 nm)|Pt(2 nm) (TeraSpinTec GmbH, Germany), where the in-plane magnetization of the ferromagnetic CoFeB is saturated by $\boldsymbol{B}_{\text{ext}}$ above 10 mT. Importantly, this sample exhibits exclusively Rashba-type SCC because Dresselhaus-type SCC is not allowed by its symmetry [10].

To characterize the ultrafast dynamics of the optically excited NiMnSb in addition to THz-emission spectroscopy [Fig. 1(a)], we probe the pump-induced changes in the MOKE and in the reflectance of the sample at wavelengths of 1550 nm, 800 nm and 515 nm. The MOKE signals exhibit dominant non-spin-related components, which mask the actual magnetization dynamics (Supplemental Note 2). The reflectance signals provide insights into the relaxation of the pump-induced change in the electron temperature.

**Typical data.**

Figs. 2 and 3 summarize THz emission data for the two samples and various magnitudes and directions $\theta$ of the external magnetic field $\boldsymbol{B}_{\text{ext}}$. Here, $\theta = \sphericalangle(\boldsymbol{B}_{\text{ext}}, [100])$ is the angle between $\boldsymbol{u}_x = [100]$ [Fig. 1(a)] of the fixed NiMnSb sample. Fig. 2(a) and 2(b) shows THz emission signals $S_x(t)$ and $S_y(t)$ from, respectively, the Rashba reference sample and NiMnSb for $\theta = 0°$ and $|\boldsymbol{B}_{\text{ext}}| = 270$ mT. We confirm that the signals are independent of the direction of the linear pump polarization and grow approximately linearly with the pump fluence (Supplemental Note 3).

Interestingly, while $S_x(t)$ is substantially smaller than $S_y(t)$ for the Rashba reference [Fig. 2(a)], $S_x(t)$ and $S_y(t)$ have comparable magnitude for NiMnSb [Fig. 2(b)]. To gain a better understanding of this behavior, it is instructive to consider the qualitative dynamics of the photocurrent sheet density $\boldsymbol{I}_c(t) = (I_{cx}, I_{cy})^{\text{T}}(t)$ [Eq. (2)]. Consequently, we display time traces of the signal vector $\boldsymbol{S}(t) = (S_x, S_y)^{\text{T}}(t)$ for the Rashba reference [Fig. 2(c)] and NiMnSb [Fig. 2(d)] and for various directions of the external field $\boldsymbol{B}_{\text{ext}}$. For the Rashba reference W|CoFeB|Pt, we observe that $\boldsymbol{S}(t)$ always moves on a line perpendicular to $\boldsymbol{B}_{\text{ext}}$ [Fig. 2(c)] and, thus, the sample magnetization $\boldsymbol{M}$, consistent with Rashba symmetry [Fig. 1(c)]. In stark contrast, for NiMnSb, the motion of $\boldsymbol{S}(t)$ is not confined to a line any more for $\theta = 0°$ and $90°$, implying that $\boldsymbol{I}_c(t)$ is not strictly perpendicular to $\boldsymbol{M}$ [Fig. 2(d)].

To quantify this observation, we determine the signal components of $\boldsymbol{S}$ parallel ($S_\parallel$) and perpendicular ($S_\perp$) to $\boldsymbol{B}_{\text{ext}}$ and $\boldsymbol{M}$. The amplitudes of $S_\parallel = \boldsymbol{S} \cdot \boldsymbol{u}_\theta$ and $S_\perp = \boldsymbol{S} \cdot \boldsymbol{u}_{\theta+90°}$ with unit vector $\boldsymbol{u}_\theta = (\cos\theta, \sin\theta)^{\text{T}}$ are displayed in Fig. 2(e,f) as a function of $\theta$ [Fig. 1(a)]. For the Rashba reference, we find that both the $S_\parallel$ and $S_\perp$ amplitudes are approximately independent of $\theta$ with $|S_\parallel| \ll |S_\perp|$. Therefore, $\boldsymbol{I}_c \perp \boldsymbol{B}_{\text{ext}}$, consistent with Fig. 2(c) and Rashba symmetry. In contrast, for NiMnSb, the $S_\parallel$ and $S_\perp$ amplitudes have components that vary according to $\cos(2\theta)$ and $\sin(2\theta)$, respectively. This more complex behavior points to physics beyond Rashba symmetry.

**Rashba- and Dresselhaus-like signal components.**

For further analysis, we tentatively assume that the suggested scenario of photoinduced spin voltage and SGE prevails [Fig. 1(a,f) and Eq. (1)]. This assumption implies $\boldsymbol{\mu}_\mathrm{s} \parallel \boldsymbol{M} \parallel \boldsymbol{B}_\mathrm{ext}$ and is consistent with the fact that the signal $S(t)$ reverses as $\boldsymbol{B}_\mathrm{ext}$ is reversed.

To simplify the SGE part of Eq. (1), we take advantage of the point symmetries of our sample, which are summarized by the group mm2. The in-plane tensor components become $-\chi_{xy} = \chi_{yx}$ and $\chi_{xx} = -\chi_{yy}$ and quantify the strength of the Rashba and Dresselhaus contribution to SCC, respectively (Supplemental Note 4). By considering the in-plane vectors $\boldsymbol{I}_\mathrm{c} = I_{\mathrm{c}x} + \mathrm{i}I_{\mathrm{c}y}$ and $\boldsymbol{B}_\mathrm{ext} = B_{\mathrm{ext}x} + \mathrm{i}B_{\mathrm{ext}y}$ as complex numbers, Eq. (1) yields the compact relationships

$$\boldsymbol{I}_\mathrm{c} = I_\mathrm{R} \mathrm{e}^{+\mathrm{i}(\theta+90°)} + I_\mathrm{D} \mathrm{e}^{-\mathrm{i}\theta}, \qquad \boldsymbol{u}_\theta = \frac{\boldsymbol{B}_\mathrm{ext}}{|\boldsymbol{B}_\mathrm{ext}|} = \mathrm{e}^{+\mathrm{i}\theta} \qquad (3)$$

with $\theta = \sphericalangle(\boldsymbol{B}_\mathrm{ext}, [100])$ [Fig. 1(a)]. Eq. (3) implies that the Rashba current $I_\mathrm{R}$ flows under an angle $\theta + 90°$, whereas the Dresselhaus current $I_\mathrm{D}$ flows under an angle $-\theta$. This directional dependence of $\boldsymbol{I}_\mathrm{c}$ vs $\theta$ is fully consistent with the Rashba [Fig. 1(c)] and Dresselhaus SGE [Fig. 1(e)] if we identify $\theta = \alpha$ in Fig. 1(a) vs Fig. 1(b-f). To compare to our experimental findings, we apply Eq. (3) to the signal components $S_\parallel = \boldsymbol{S} \cdot \boldsymbol{u}_\theta$ and $S_\perp = \boldsymbol{S} \cdot \boldsymbol{u}_{\theta+90°}$. We find that $S_\parallel \propto \mathrm{Re}(\boldsymbol{I}_\mathrm{c} \mathrm{e}^{-\mathrm{i}\theta})$ is a superposition of $I_\mathrm{R} \cos(90°) = 0$ and $I_\mathrm{D} \cos(2\theta)$, whereas $S_\perp \propto \mathrm{Re}(\boldsymbol{I}_\mathrm{c} \mathrm{e}^{-\mathrm{i}(\theta+90°)})$ is a linear combination of $I_\mathrm{R} \sin(90°) = I_\mathrm{R}$ and $I_\mathrm{D} \sin(2\theta)$. This conclusion agrees with the $\theta$-dependence of the measured $S_\parallel$ and $S_\perp$ amplitudes [Fig. 2(e,f)].

To demonstrate the consistence of Eq. (3) with our data even more directly and to extract the coefficients $I_\mathrm{R}$ and $I_\mathrm{D}$, we Fourier-transform $\boldsymbol{I}_\mathrm{c}$ of Eq. (3) with respect to $\theta$. The Fourier-transformed $\boldsymbol{I}_\mathrm{c}$ has a peak at frequency (annular multiplicity) $+1$ and $-1$, whose respective amplitude scales with $\mathrm{e}^{+\mathrm{i}90°} I_\mathrm{R} = \mathrm{i}I_\mathrm{R}$ and $I_\mathrm{D}$. For comparison, Fig. 3(a,b) displays the $\theta$-related Fourier amplitude of the measured signals $\boldsymbol{S} = S_x + \mathrm{i}S_y$. The signal from the Rashba reference sample is dominated by a component $\propto \mathrm{e}^{+\mathrm{i}\theta}$ with imaginary amplitude [Fig. 3(a)], consistent with the Rashba term ($\propto \mathrm{i}I_\mathrm{R} \mathrm{e}^{+\mathrm{i}\theta}$) in Eq. (3). In contrast, the signal from NiMnSb contains a Rashba component as well, but also another component $\propto \mathrm{e}^{-\mathrm{i}\theta}$ with a real-valued amplitude [Fig. 3(b)], as expected from a Dresselhaus term ($\propto I_\mathrm{D} \mathrm{e}^{-\mathrm{i}\theta}$) in Eq. (3).

We use the Fourier approach of Figs. 3(a,b) to extract Rashba ($\propto I_\mathrm{R}$) and Dresselhaus ($\propto I_\mathrm{D}$) signal components from the Rashba reference and NiMnSb. The resulting 4 waveforms are displayed in Fig. 3(c) vs time $t$. As expected, the Rashba reference sample delivers a negligible Dresselhaus signal. In contrast, NiMnSb exhibits Rashba and Dresselhaus components of comparable amplitude. Note that their dynamics is substantially different, suggesting they have a different origin.

Finally, Fig. 3(d) shows the amplitude of the Rashba and Dresselhaus signal components from NiMnSb as a function of the magnitude $|\boldsymbol{B}_\mathrm{ext}|$ of the external magnetic field. Interestingly, the Dresselhaus component is approximately independent of $|\boldsymbol{B}_\mathrm{ext}|$, whereas the Rashba component increases approximately linearly with $|\boldsymbol{B}_\text{ext}|$. Consequently, we ascribe the Rashba component to an out-of-plane photocurrent that is converted into an in-plane charge current by the ordinary Hall effect. The latter has Rashba symmetry but is not included in Eq. (1) because it is not related to SCC.

In contrast, the Dresselhaus component is independent of $|\boldsymbol{B}_\text{ext}|$ [Fig. 3(d)]. As the covered values of $|\boldsymbol{B}_\text{ext}|$ are larger than the saturation magnetic field of our sample (50 mT, Supplemental Note 2), we conclude that the Dresselhaus component scales with the NiMnSb magnetization $|\boldsymbol{M}|$. According to Eq. (1), this signal may not only arise from $\boldsymbol{\mu}_\text{s}$ and the SGE but also a pump-induced spin current $j_\text{s}\boldsymbol{\sigma}\otimes\boldsymbol{u}_z$ with polarization $\boldsymbol{\sigma}=\boldsymbol{M}/|\boldsymbol{M}|$ that flows along $\boldsymbol{u}_z$ from NiMnSb to Ru and undergoes SCC by a Dresselhaus-type ISHE in NiMnSb [Fig. 1(a)]. The latter scenario makes a minor contribution, as indicated by measurements on NiMnSb|MgO stacks where electron transport from NiMnSb to the insulating MgO is inhibited (Supplemental Note 3). We conclude that the Dresselhaus signal [Fig. 3(c)] is fully compatible with an optically induced spin accumulation $\boldsymbol{\mu}_\text{s}$ [Fig. 1(f)] and its conversion into a charge current by a SGE with Dresselhaus symmetry in the NiMnSb bulk [Fig. 1(e)].

We note that the observed dependence of the THz signal on the angle $\theta=\sphericalangle(\boldsymbol{M},[100])$ of the magnetization $\boldsymbol{M}$ [Figs. 1(a) and 3(a,b)] implies that the SGE tensor $\chi$ [Eq. (1)] is largely independent of $\boldsymbol{M}$. Conversely, a pronounced $\boldsymbol{M}$-dependence of $\chi$ would result in a more complex dependence on $\alpha=\theta$ than in Fig. 1(c,e), which is not observed. Previous work on spin-orbit torque in NiMnSb arrived at an analogous conclusion for the inverse SGE [4,39]. We conclude that the subsystem (1) of electronic states carrying the NiMnSb magnetization $\boldsymbol{M}$ and the subsystem (2) of electronic states providing the Dresselhaus SGE do not coincide and are only weakly coupled, as schematically shown in Fig. 4(a).

**Dresselhaus current dynamics.**

To gain more insight into the generation of the Dresselhaus-type charge current $I_\text{D}$, we extract its dynamics vs time $t$ from the respective THz signal in Fig. 3(c) [40]. The resulting $I_\text{D}(t)$ [Fig. 4(b)] increases on a time scale of 50 fs, which is most likely given by the time resolution of the experiment, and relaxes on a scale of 0.2 ps.

It is instructive to compare $I_\text{D}(t)$ to the pump-induced change $\Delta R(t)$ in the sample reflectance at a probe wavelength of 515 nm [Fig. 4(c)]. Interestingly, once $\Delta R(t)$ has reached its peak value, it decays with a similar time constant as $I_\text{D}(t)$. In metals, the relaxation dynamics of $\Delta R(t)$ typically reports on the cooling dynamics of the electrons by the phonons [40,41]. Therefore, the similar relaxation dynamics of $I_\text{D}(t)$ [Fig. 4(b)] and $\Delta R(t)$ [Fig. 4(c)] indicates that the Dresselhaus current largely follows the dynamics of the excess energy of the optically excited electrons in NiMnSb.

This observation suggests the following scenario of our experiment [Fig. 4(a)]. The pump pulse heats the electrons of NiMnSb and, consequently, induces an excess of spin $\boldsymbol{\mu}_\text{s}^M$ in electron subsystem (1) [Fig. 1(f)]. In turn, $\boldsymbol{\mu}_\text{s}^M$ triggers transfer of spin angular momentum to the crystal lattice [26], but also to electron subsystem (2) [Fig. 4(a)] where it induces a spin excess $\boldsymbol{\mu}_s$ that gives rise to the Dresselhaus-type SGE [Fig. 1(e)].

Note that spin relaxation in electron subsystem (2) proceeds on the time scale of momentum relaxation, owing to the Dresselhaus-type locking of spin and momentum [Fig. 1(d)]. The recent estimate of the momentum relaxation times of 7 fs [39] is consistent with values found for thin metal films [17,42,43]. Consequently, $\boldsymbol{\mu}_s$ follows $\boldsymbol{\mu}_s^M$ instantaneously within the time resolution of our experiment, and we have

$$I_D(t) \propto \boldsymbol{\mu}_s(t) \propto \boldsymbol{\mu}_s^M(t) \tag{4}$$

to good approximation. In other words, the Dresselhaus current $I_D(t)$ [Fig. 4(b)] monitors the dynamics of the spin excess in both subsystems (1) and (2).

In ferromagnetic metals, such as Ni and CoFeB, the dynamics of $\boldsymbol{\mu}_s(t)$ is given by a competition between (i) the pump-induced change in the electronic temperature through the mechanism depicted in Fig. 1(e) and (ii) a possible transfer of spin angular momentum to the crystal lattice [26] (Supplemental Note 5). Process (i) and (ii) is quantified by the time constant $\tau_{ep}$ and $\tau_{sl}$ of electron-phonon and spin-lattice equilibration, respectively. We obtain the value of $\tau_{ep}$ from fitting the measured reflectance $\Delta R(t)$ by an exponential decay with time constant $\tau_{ep}$ plus offset [Eq. (16)] [40], which yields $\tau_{ep} = 0.25$ ps [Fig. 4(c)]. Finally, we fit $I_D(t)$ by a sum of 2 exponentials with time constants $\tau_{sl}$ and the known $\tau_{ep}$ [Eq. (17)] [26] and obtain very good agreement of experiment and theory for $\tau_{sl} \sim 1$ ps.

We conclude that, in NiMnSb, spin-lattice equilibration ($\tau_{sl}$) is substantially slower than electron-phonon equilibration ($\tau_{ep}$). The relatively long $\tau_{sl}$ is in line with the half-metal characteristics of NiMnSb that strongly suppresses spin-flip scattering [44,45]. Because $\tau_{sl} \gg \tau_{ep}$, the decay of $I_D(t) \propto \boldsymbol{\mu}_s(t)$ is dominated by electron-phonon relaxation, consistent with the similar relaxation dynamics of $I_D(t)$ [Fig. 4(b)] and $\Delta R(t)$ [Fig. 4(c)]. This behavior is opposite to metallic ferromagnets, such as Ni and CoFeB, where the relaxation of $\boldsymbol{\mu}_s$ is dominated by $\tau_{sl} \ll \tau_{ep}$ [26].

**In conclusion**, we observe ultrafast photocurrents in the metallic ferromagnetic Heusler compound NiMnSb. They are consistent with a Dresselhaus-type SGE in the inversion-asymmetric bulk of the material. The response time of this SCC mechanism is given by the electron momentum relaxation time of ~10 fs and, thus, genuinely ultrafast. Our results indicate that THz-emission-based photocurrent methodology is a powerful tool to gain insight into the so far underexplored class of inversion-asymmetric ferromagnets. In contrast to ferromagnet|heavy-metal heterostructures, where the photocurrent is generated only in the interface regions, the photocurrent in NiMnSb is expected to flow throughout the bulk. By exploring this scaling effect of inversion-asymmetric ferromagnets, we expect to discover new powerful emitters [10–12] of ultrabroadband THz electromagnetic pulses. Finally, by reversing the concept of spintronic THz emission [Fig. 1(a)], promising perspectives for detecting THz fields [17] and ultrafast magnetization switching by bulk spin-orbit torques [18–20] arise.